\documentclass[prd,twocolumn,superscriptaddress]{revtex4-1}   
\usepackage{graphicx}
\usepackage{multirow}
\usepackage{gensymb}
\usepackage{amsmath}
\usepackage{bm}
\usepackage{natbib}
\usepackage{hyperref}
\usepackage{color}  

\newcommand\aj{{AJ}}
\newcommand\araa{{ARA\&A}}
\renewcommand\apj{{ApJ}}
\newcommand\apjl{{ApJL}}
\newcommand\apjs{{ApJS}}
\newcommand\aap{{A\&A}}
\newcommand\aapr{{A\&A~Rev.}}
\newcommand\mnras{{MNRAS}}
\renewcommand\prd{{Phys.~Rev.~D}}
\renewcommand\prl{{Phys.~Rev.~Lett.}}
\newcommand\pasp{{PASP}}
\renewcommand\nat{{Nature}}
\newcommand\nar{{New A Rev.}}

\def\mone{{\rm M}_1}
\def\mtwo{{\rm M}_2}

\def\nintr{{\rm N}_0}
\def\kintr{{\mathcal N}_0}
\def\ngttwo{{\rm N}_{>2}}
\def\ngtfive{{\rm N}_{>5}}
\def\ngtten{{\rm N}_{>10}}
\def\ngts{{\rm N}_{>s}}
\def\kgttwo{{\mathcal N}_{>2}}
\def\kgtfive{{\mathcal N}_{>5}}
\def\kgtten{{\mathcal N}_{>10}}
\def\kgts{{\mathcal N}_{>s}}
\def\nsamp{{\rm N}_{\rm samp}}
\def\mecc{\overline e}
\def\kapone{\kappa_1}
\def\dmax{D_{\rm max}}
\def\zmin{Z_{\rm min}}
\def\zmax{Z_{\rm max}}
\def\zf{z_{\rm f}}
\def\tf{t_{\rm f}}
\def\zd{z_{\rm D}}
\def\tevnt{t_{\rm event}}
\def\tdelay{t_{\rm delay}}
\def\tld{t_{\rm lD}}
\def\tobs{t_{\rm obs}}
\def\delobs{\Delta t_{\rm obs}}
\def\tlbf{t_{\rm lb,f}}
\def\hn{h_{\mathcal N}}
\def\sn{S_{\mathcal N}}
\def\pn{P_{\mathcal N}}
\def\respf{\mathcal R}
\def\sgal{S_{\rm g}}

\newcommand{\Ms}{\ensuremath{M_{\odot}}}
\newcommand{\Zs}{\ensuremath{Z_{\odot}}}
\newcommand{\eg}{{\it e.g.}}

\newcommand{\ie}{{\it i.e.}}

\newcommand{\beq}{\begin{equation}}
\newcommand{\eeq}{\end{equation}}
\newcommand{\mtot}{\ensuremath{M_{\rm tot}}}

\newcommand{\mch}{\ensuremath{M_{\rm ch}}}
\newcommand{\mchz}{\ensuremath{M_{\rm ch,z}}}

\newcommand{\kmps}{\ensuremath{{\rm~km~s}^{-1}}}

\newcommand{\thub}{\ensuremath{t_{\rm Hubble}}}

\newcommand{\rhocl}{\ensuremath{\rho_{\rm cl}}}
\newcommand{\mcl}{\ensuremath{M_{cl}}}
\newcommand{\rh}{\ensuremath{r_h}}

\newcommand{\nbseven}{{\tt NBODY7}}

\newcommand{\bse}{{\tt BSE}}

\newcommand{\archain}{{\tt ARCHAIN}}

\newcommand{\tlisa}{\ensuremath{T_{\rm LISA}}}

\newcommand{\fgwp}{\ensuremath{f_{\rm GWp}}}
\newcommand{\fgwpz}{\ensuremath{f_{\rm GWp,z}}}
\newcommand{\fdotgwp}{\ensuremath{\dot{f}_{\rm GWp}}}
\newcommand{\fdotgwpz}{\ensuremath{\dot{f}_{\rm GWp,z}}}
\newcommand{\fk}{f_{\rm K}}

\newcommand{\hc}{\ensuremath{h_{\rm c}}}
\newcommand{\sbyn}{\ensuremath{\left(\frac{\rm S}{\rm N}\right)}}

\begin{document}

\title[LISA sources from young clusters]
{LISA sources from young massive and open stellar clusters}

\author{Sambaran Banerjee}
\email{sambaran@astro.uni-bonn.de (SB)}
\affiliation{Helmholtz-Instituts f\"ur Strahlen- und Kernphysik (HISKP),
Nussallee 14-16, D-53115 Bonn, Germany}
\affiliation{Argelander-Institut f\"ur Astronomie (AIfA), Auf dem H\"ugel 71,
D-53121, Bonn, Germany}

\date{\today}

\begin{abstract}
I study the potential role of young massive star clusters (YMCs)
and open star clusters (OCs) in
assembling stellar-mass binary black holes (BBHs) which would be
detectable as persistent gravitational-wave (GW)
sources by the forthcoming, space-based Laser Interferometer Space Antenna (LISA).
The energetic dynamical interactions inside star clusters make them factories of
assembling BBHs and other types of double-compact binaries that undergo
general-relativistic (GR) inspiral and merger. The initial phase
of such inspirals would, typically, sweep through the LISA GW band.
This fabricates a unique opportunity to probe into
the early in-spiralling phases of merging BBHs, that would provide insights
into their formation mechanisms. Here, such LISA sources are studied
from a set of evolutionary models of star clusters with masses ranging over $10^4\Ms-10^5\Ms$
that represent YMCs and intermediate-aged OCs in
metal-rich and metal-poor environments of the Local Universe.
These models are evolved with long-term, direct, relativistic many-body computations
incorporating state-of-the-art stellar-evolutionary and remnant-formation models.
Based on models of Local Universe constructed with such model clusters, it is
shown that YMCs and intermediate-aged OCs would yield several 10s to 100s of LISA BBH sources
at the current cosmic epoch with GW frequency within $10^{-3}{\rm~Hz} - 10^{-1}{\rm~Hz}$
and signal-to-noise-ratio (S/N) $>5$, assuming a mission lifetime of 5 or 10 years.
Such LISA BBHs would have a bimodal distribution in total mass, be generally eccentric ($\lesssim0.7$), 
and typically have similar component masses although mass-asymmetric systems are
possible.
Intrinsically, there would be 1000s of present-day, LISA-detectable BBHs from YMCs and OCs.
That way, YMCs and OCs would provide a significant and the dominant contribution to
the stellar-mass BBH population detectable by LISA. A small fraction, $<5$\%, of these
BBHs would undergo GR inspiral to make it to LIGO-Virgo GW frequency band and merge,
within the mission timespan; $<15$\% would do so within twice the timespan.
LISA BBH source counts for a range of S/N, normalized w.r.t. the local cluster density,
are provided. Drawbacks in the present approach and future improvements are discussed. 
\end{abstract}

\maketitle

\section{Introduction}\label{intro}

Following the recent, above-the-expectation success of the LISA Pathfinder mission \citep{LPF},
Laser Interferometer Space Antenna (LISA; also, eLISA) has now been approved as an L3 mission
by the European Space Agency \citep{eLISA}. LISA is a proposed space-borne, dual-arm
interferometer gravitational wave (hereafter GW) detector with an arm length of $2.5\times10^6$ km.
Such arm length makes the instrument sensitive to GWs of much lower frequencies,
$\sim 10^{-5}-10^{-1} {\rm~Hz}$ \citep{eLISA},
compared to its ground-based counterparts \citep{Abadie_2010,2016PhRvL.116f1102A}.
LISA will thus potentially observe a wide-variety of low-frequency GW events such as mergers of
binary supermassive black holes, inspiral and mergers of binary intermediate-mass black holes,  
intermediate mass ratio and
extreme mass ratio inspirals involving supermassive and intermediate-mass black holes,
Galactic binary white dwarfs and binary stars, and stellar-remnant binary black holes (hereafter BBH)
\footnote{In this work, `BBH' will imply binary black holes composed of
stellar-remnant/stellar-mass black holes.}
and other double-compact binaries
in the Local Universe \citep{eLISA,AmaroSeoane_2007,Ruiter_2010,Holley_2015,Sesana_2016}.

Over their first (O1), second (O2), and third (O3) observing runs, the LIGO-Virgo collaboration (hereafter LVC)
has identified 67 compact-binary merger events, which are predominantly BBH merger candidates
but also contain binary neutron star (hereafter BNS) and neutron star-black hole (hereafter NSBH) merger    
events. Among these, the parameter estimations of 11 BBH mergers and 2 BNS mergers have
so far been published (\cite{Abbott_GWTC1,LSC_GW190425,LSC_GW190412},
\url{https://gracedb.ligo.org/superevents/public/O3/}). However, various theories
leading to compact-binary mergers and their observed properties (see \cite{Benacquista_2013} for a review)
still remain largely degenerate.

One of the main reasons for this degeneracy is the fact that most compact binaries ``forget''
their orbital parameters at formation and hence the imprints of their formation
mechanisms, by shrinking to a large extent and becoming practically circular by
the time they spiral in, via GW radiation, up to the LIGO-Virgo GW frequency band
($\sim 10 - 1000$ Hz). By probing BBHs and other double-compact binaries at GW frequencies
that are lower by a few orders of magnitude, LISA has the potential to identity imprints
of such systems' formation mechanisms. In that sense, identification of
BBHs and other double-compact binaries by LISA and as well by other proposed,
deci-Hertz-range, space-based GW interferometers such as DECIGO \citep{Kawamura_2008,ArcaSedda_2019}
and Tian Qin \citep{Luo_2016,Liu_2020} would be complementary to the ground-based
general-relativistic (hereafter GR) merger detections.

In particular, it can generally be expected that BBHs assembled via dynamical
interactions in stellar clusters would be eccentric. Dynamically-formed BBHs that
merge within a Hubble time would exhibit relics of this eccentricity in the LISA 
frequency band \citep{Nishizawa_2016,Nishizawa_2017}, on their way
to the merger via post-Newtonian (hereafter PN) inspiral. Detailed and self-consistent
direct N-body and Monte Carlo simulations of young, open, and globular clusters
indeed support this \citep{Banerjee_2017b,Kremer_2019,Banerjee_2020c}. 
In contrast, isolated binary evolution can be expected to produce predominantly 
circular BBHs in the LISA band. This is because to place a BBH, derived from
massive stellar binary evolution, in the LISA band, the binary must
go through a common-envelope (CE) phase \citep{Ivanova_2013} so that it shrinks sufficiently
\citep{Belczynski_2016,Mandel_2017,Stevenson_2017,Giacobbo_2018,Baibhav_2019},
which process would also circularize them.
Except for the least massive merging BBHs produced in this way (which would also 
have the least chances to be visible by LISA), that may become eccentric at the
beginning of their GR-inspiral due to BHs' natal kick \citep{Banerjee_2020} (especially, that of the
later-born BH), the BH members would form via direct collapse without
any natal kick, preserving the circular binary orbit
\footnote{If the natal kick of stellar remnants is predominantly due to asymmetric
emission of neutrinos \citep{Fuller_2003,Fryer_2006},
then direct collapse BHs would also receive significant natal
kicks \citep{Banerjee_2020}. In that case, finding out how orbital characteristics of field BBHs in the LISA band
would compare with dynamically-assembled BBHs requires detailed modelling of
neutrino-driven kick in population synthesis of massive binaries.}.

Note that the typical timescale of PN orbital evolution of BBHs 
in the LISA band is $\sim 0.1$ Myr although, depending on the BBH's
orbital configuration, it can be as small as $\sim 10$ yr \citep{Banerjee_2020c}.
Therefore, BBHs in the LISA band are persistent or semi-persistent GW sources.
In contrast, they are transient GW sources in the LIGO-Virgo band, the inspiral
timescale being $\lesssim$ a minute.

The contribution of BBH LISA sources from globular clusters (hereafter GC) and nuclear
clusters (hereafter NSC) in the Local Universe, due to dynamical processes in such clusters,
has recently been studied \citep{Samsing_2018a,Kremer_2019,Hoang_2019}.
This study investigates BBH LISA sources from young massive clusters (hereafter YMC)
and open clusters (hereafter OC) which aspect
is rather unexplored to date. To that end, the set of theoretical cluster evolutionary models
as described in \citep{Banerjee_2020c} is utilized. The structure and stellar
composition of these cluster models are consistent with those observed in YMCs and OCs in the
Milky Way and the Local Group. The models are evolved with state-of-the-art
PN direct N-body integration, incorporating up-to-date supernova (hereafter SN) and
stellar remnant formation models.

In Sec.~\ref{nbcomp}, the N-body evolutionary models of star clusters 
and BBH inspirals from them are summarized.
Sec.~\ref{lisacount} discusses the method of constructing models of Local Universe
with these cluster models and obtaining present-day LISA BBH sources
from them.
Sec.~\ref{lisasrc} estimates the LISA BBH source counts and the sources'
properties.
Sec.~\ref{discuss} summarizes and discusses the present results, their caveats,
and suggests upcoming improvements. 

\section{Computations}\label{comp}

In this section, the approach to determine LISA source count and 
properties, based on model cluster evolution, is described. 

\subsection{Post-Newtonian, many-body cluster-evolutionary models}\label{nbcomp}

In this work, the 65 N-body evolutionary models of star clusters,
as described in \citep{Banerjee_2020c}, are utilized. The model clusters,
initially, possess a Plummer density profile \citep{Plummer_1911} for
the spatial distribution of all constituent stars,
are in virial equilibrium \citep{Spitzer_1987,Heggie_2003},
have masses $10^4\Ms\leq\mcl(0)\leq10^5\Ms$, and have half-mass radii 
$1{\rm~pc}\leq\rh(0)\leq3{\rm~pc}$. They range over 
$0.0001\leq Z \leq0.02$ in metallicity and are subjected to a solar-neighborhood-like
external galactic field. The initial models are composed of zero-age-main-sequence
(hereafter ZAMS) stars with masses over $0.08\Ms-150.0\Ms$ and distributed according
to the standard initial mass function (hereafter IMF). About half of these models
have a primordial-binary population (overall initial binary fraction
$\approx5$\% or 10\%) where all the O-type stars (\ie, stars
with ZAMS mass down to $16\Ms$) are paired among themselves
with an observationally-motivated distribution of massive stellar binaries
\citep{Sana_2011,Moe_2017}. Although idealistic, such cluster
parameters and stellar compositions are consistent with those observed
in YMCs and medium-mass OCs that continue to form and dissolve in
the Milky Way and other Local-Group galaxies.

These model clusters are evolved using $\nbseven$,
a state-of-the-art PN direct N-body integrator
\citep{Aarseth_2003,Aarseth_2012,Nitadori_2012}, that couples with the
semi-analytical (or population synthesis) stellar and binary-evolutionary model
$\bse$ \citep{Hurley_2000,Hurley_2002}. The integrated $\bse$ is made up-to-date
\citep{Banerjee_2020} in regards to prescriptions of stellar wind mass loss \citep{Belczynski_2010},
formation of stellar-remnant neutron stars (hereafter NS) and black holes (hereafter BH) 
by incorporating the `rapid' and `delayed' SN models \citep{Fryer_2012} and
pulsation pair-instability (PPSN) and pair-instability (PSN) SN \citep{Belczynski_2016a}.  
The SN remnants receive natal velocity kicks that are scaled down from a Maxwellian distribution
with dispersion equal to the velocity dispersion of single NSs in the Galactic field
\citep[$\approx265\kmps$;][]{Hobbs_2005}. The scale-down is applied based on
material fallback onto the proto-remnant in the SN \citep{Fryer_2012} and according to the
conservation of linear momentum (popularly referred to as the `momentum conserving natal kick'
\cite{Belczynski_2008,Giacobbo_2018}). This slow-down procedure allows the
natal kicks of $\gtrsim10\Ms$ BHs \citep{Banerjee_2020} to be less than the
parent clusters' escape speeds which BHs are, therefore, retained back in the clusters right after
their birth.

The PN treatment of $\nbseven$ is handled by $\archain$ sub-integrator \citep{Mikkola_1999,Mikkola_2008}
that applies PN corrections (up to PN-3.5) to a binary with an
NS or a BH component that either is by itself gravitationally bound
to the cluster or is a part of an in-cluster triple or higher-order subsystem.
The (regularized) PN orbital integration of the binary
takes into account perturbations from the outer members (if part of a subsystem)
until the subsystem is resolved, via either the binary's GR inspiral and coalescence
or the disintegration of the subsystem. This allows in-cluster
GR mergers driven by the Kozai-Lidov (hereafter KL)
mechanism \citep{Kozai_1962,Lithwick_2011,Katz_2011} or
chaotic triple (or higher order) interactions \citep[\eg,][]{Antonini_2016b,Samsing_2018a}.
Apart from such in-cluster mergers, which comprise the majority of the
GR mergers from these model clusters, a fraction of the
double-compact binaries ejected dynamically from the clusters
would also undergo PN inspiral and merger within
a Hubble time
\citep[\eg,][]{PortegiesZwart_2000,Banerjee_2010,Rodriguez_2015,Kumamoto_2019,DiCarlo_2019}.
As demonstrated in \citep[][see also \cite{Kremer_2019}]{Banerjee_2017b,Banerjee_2020c},
the vast majority of such in-cluster and ejected dynamically-driven inspirals, most
of which are BBH inspirals,
initiate with peak GW frequency lying within or below the LISA band.
Although most of these inspirals begin with very high eccentricity,
they circularize via GR inspiral \citep{Peters_1964} to become
moderately eccentric ($\lesssim0.7$) within the LISA band, and be visible by the instrument
\citep{Nishizawa_2016,Nishizawa_2017,Chen_2017}.

The stellar-remnant BHs are assigned spins at birth based on hydrodynamic
models of fast-rotating massive single stars \citep{Belczynski_2020} 
which are utilized in assigning numerical-relativity-based GR merger recoil kicks and
final spins \citep{Baker_2008,Rezzolla_2008,vanMeter_2010} of the in-cluster BBH mergers. 
However, the $\archain$ PN integrations are themselves performed  
assuming non-spinning members for the ease of computing; this
simplification is not critical for LISA GW frequencies since spin-orbit precession
and the corresponding modification of orbital-evolutionary time
would be mild over such frequencies. In the computed models of \citep{Banerjee_2020c},
the majority of the BBH mergers have primaries $\mone\lesssim40\Ms$. However, although rarely,
$\mone$ reaches up to $\approx100\Ms$ (total mass up to $\approx140\Ms$),
due to the occurrence of second-generation BBH mergers \citep{Gerosa_2017,Rodriguez_2018}
or BBH mergers involving a BH that has previously gained mass via
merging with a regular star (forming a BH Thorne-Zytkow object) \citep{Banerjee_2020c}.
Further details of these computed star cluster models are given in \cite{Banerjee_2020c}
and the stellar and binary-evolutionary schemes used in these models are
further elaborated in \cite{Banerjee_2020}.

Note that the present cluster models initiate with a Plummer density profile
in virial equilibrium.
However, the initial density profile is unlikely to significantly influence the GR
inspiral events (their number, rate, and properties) from the clusters. This
is because the dynamically-triggered GR inspirals take place
due to the close encounters in the `BH core' which is formed after the BHs
(that are retained in the cluster after their birth; see above) segregate in the innermost
region of the cluster \citep{Banerjee_2010,Morscher_2015},
in $\sim100$ Myr for the masses and sizes of the present
models \citep{Spitzer_1987}. The properties of this BH core (or BH sub-cluster)
and hence of the BBHs formed inside it depend mostly
on the bulk properties of the whole star cluster such as its mass and virial radius
\citep{Henon_1975,Breen_2013,Antonini_2020}.
Note further that, in reality, the (proto-)clusters would initially have been sub-virial and contained substructures,
or, alternatively, would have gone through a super-virial phase due to residual gas expulsion, 
as suggested by observations of molecular clouds, young stellar nurseries,
and young clusters \citep{Longmore_2014,Feigelson_2018,Banerjee_2018b}.
However, as long as a cluster survives such an initial `violent-relaxation' phase
(so that it becomes a `fully formed cluster' as modelled here; see also Sec.~\ref{discuss}),
the substructures would be washed out
and the cluster would become (near-)spherical and virialized in a few dynamical (or free-fall) 
times, typically in $\sim$ Myr \citep{Marks_2012,Marks_2012b,Banerjee_2015b,Brinkmann_2017,Shukirgaliyev_2017},
\ie, much earlier than the `BH core' formation.

The initially $\approx100\%$ primordial binary fraction among the
BH-progenitor O-type stars (see above) moderately
affects, through binary evolution \citep{Spera_2019,DiCarlo_2019,Banerjee_2020,Banerjee_2020c},
the mass distribution of the BHs retained in the clusters. However, since inside
a dynamically-active BH core (\ie, which core is efficiently producing BBHs and their inspiral-mergers
through dynamical interactions)
the BHs interact mostly among themselves and rarely with
the regular stars and their binaries \citep{Chatterjee_2017a,Kremer_2017,Banerjee_2018},
the primordial binary fraction among the lower-mass stars
($\approx5$\% or 10\% in the present models; see above) is unlikely to largely influence
the BBH production (the dynamical heating from the stellar binaries may mildly affect
the structure of the cluster and hence of the BH sub-system).
In some computed clusters, an alternative, `collapse-asymmetry-driven' natal kick
model \citep{Burrows_1996,Fryer_2004,Meakin_2006,Meakin_2007}
is applied instead of the standard momentum-conserving kick (see above).
The collapse-asymmetry-driven kick model, in addition to incorporating slow   
down due to SN material fallback, applies additional slowing down mechanism (with recipes based
on numerical computations of stellar collapse) for NSs and low mass BHs ($\lesssim10\Ms$)
arising from washing-out of asymmetries in the pre-SN star due to convection;
see \citep{Banerjee_2020} and references therein for further detail.
Since young and open clusters, which are of age $\lesssim$ Gyr, are considered
here (see Sec.~\ref{lisacount}), the least massive BHs would mostly be dynamically
inert over the clusters' age range \citep[\eg,][]{Kremer_2020}.
Therefore, the use of this alternative natal kick recipe in a few models
is unlikely to have a substantial influence on the overall BBH/inspiral-merger yield from these models.

In summary, current specifics of the initial cluster models or their potential
alternatives (\eg, use of initially King or fractal profiles instead of Plummer,
higher primordial binary fraction among non-BH-progenitor stars) would, at best, have
order-unity influence on the BBH production and their GR-inspiral and merger events
from the models. The initial mass and size ranges considered in these modes are consistent
with those observed for young massive clusters in the Local Universe \citep{PortegiesZwart_2010}.
Alternative model ingredients will be explored in a future study.

\subsection{Present-day LISA sources from computed cluster models}\label{lisacount}

A `sample Local Universe' is constructed out of $\nsamp$ model clusters by
placing each cluster at a random
comoving distance, $D$, within a spherical volume of $\dmax=1500$ Mpc, centered
around the detector.
The value of $\dmax$ is set based on the fact that at this distance the brightest LISA
sources from the computed models still project characteristic strain
marginally above LISA's design noise floor (see below). In other words, $\dmax$ is the
limit of visibility of BBH sources, as of the present computed models.

For each cluster at a chosen distance $D$, a model of mass, $\mcl(0)$, is selected
from the set of computed cluster models,
with probability $\propto\mcl(0)^{-2}$ over its range of $10^4\Ms-10^5\Ms$ in the set.
Such a mass distribution is observed for
newborn and young clusters of a wide mass range in the Milky Way and nearby galaxies
\citep{Lada_2003,Gieles_2006a,Gieles_2006b,Larsen_2009}. The model's size
is selected uniformly over its range in the
computed set, \ie, $1{\rm~pc}\leq\rh(0)\leq3{\rm~pc}$ and
its metallicity is chosen uniformly over $\zmin\leq Z \leq\zmax$.
Two metallicity ranges are considered: that of the entire model set $(\zmin,\zmax)=(0.0001,0.02)$
ranging from very metal-poor environments up to the solar enrichment and
$(\zmin,\zmax)=(0.005,0.02)$ comprising only metal-rich systems (down to
$0.25\Zs$). Note that most of the models in the wider $Z$-range case 
have $Z\geq0.001$ \citep[see][]{Banerjee_2020c} as consistent with
the most metal-poor galaxies observed in the Local Universe \citep{Hsyu_2018}.
The choice of two $Z$ ranges allows studying the impact of metallicity
on LISA source counts and properties. As shown in Table~\ref{tab_nlisa},
Local-Universe samples of $\nsamp\sim10^4$ are considered which sample sizes
provide a fair balance between the computing time required for
extracting the present-day LISA sources (see below) and statistics.

Each cluster in a sample is assigned a formation redshift, $\zf$, that
corresponds to an age, $\tf$, of the Universe. A GR inspiral in the LISA frequency band
occurs from this cluster
(the vast majority of the inspirals are of BBH; see \cite{Banerjee_2020c})
after a delay time, $\tdelay$, from the formation
when the age of the Universe is $\tevnt$, \ie,
\beq
\tevnt = \tf + \tdelay. 
\label{eq:tevnt}
\eeq
If the light travel time from the cluster's distance, $D$, is $\tld$, then
the age of the Universe is
\beq
\tobs = \tevnt + \tld
\label{eq:tobs}
\eeq
when the (redshifted) GW signal reaches the detector. The formation epoch, $\zf$,
of a cluster is assigned according to the probability distribution given
by the cosmic star formation history (hereafter SFH), namely \citep{Madau_2014},
\beq
\psi(z) = 0.015\frac{(1+z)^{2.7}}{1+[(1+z)/2.9]^{5.6}} \Ms{\rm~yr}^{-1}{\rm~Mpc}^{-3}.
\label{eq:mdsfr}
\eeq
A detected signal is considered `recent' if
\beq
\thub-\delobs \leq \tobs \leq \thub+\delobs
\label{eq:delobs}
\eeq
where $\thub$ is the present age of the Universe (the Hubble time) and $\delobs$ is taken
to be $\delobs=0.1{\rm~Gyr}$. $\delobs$ serves as an uncertainty in the cluster formation
epoch; with the above choice it is well within the typical epoch uncertainties in the
observed SFH data \citep{Madau_2014}. In this work, the contribution of
LISA sources from young and open clusters are considered which is why
the formation epoch is restricted to relatively recent times, namely,
$0.0\leq\zf\leq0.5$ that corresponds to formation look-back times within
$0.0{\rm~Gyr}\leq\tlbf\lesssim5{\rm~Gyr}$.

\begin{table*}
	\caption{LISA source counts based on representative samples of the Local Universe ($D\leq1500$ Mpc), as
	constructed with the computed model clusters (Secs.~\ref{nbcomp} and \ref{lisacount}). The columns from
	left to right are as follows:
	Col. 1: metallicity ($Z$) range of the model clusters in a sample,
	Col. 2: lifetime of the LISA mission, $\tlisa$,
	Col. 3: number of clusters, $\nsamp$, in the sample,
	Col. 4: intrinsic number of LISA sources, $\nintr$, at the present cosmic age (within $\thub\pm0.1$ Gyr)
	from the sample,
	Col. 5: inferred intrinsic number of LISA sources within $\tlisa$, $\kintr$, 
	scaled by the cluster density (in ${\rm~Mpc}^{-3}$) of the Local Universe, $\rhocl$, 
	Cols. 6, 8, 10: numbers of LISA sources from the sample, $\ngttwo$, $\ngtfive$, and $\ngtten$,
	with S/N $\geq2$, $\geq5$, and $\geq10$ respectively, at the present cosmic age,
	Cols. 7, 9, 11: inferred numbers of LISA sources within $\tlisa$, $\kgttwo$, $\kgtfive$, and $\kgtten$,
	with S/N $\geq2$, $\geq5$, and $\geq10$ respectively,
	scaled by the cluster density of the Local Universe (Eqn.~\ref{eq:nlisasc}).
	\vspace{0.3 cm}
	}
\label{tab_nlisa}
\centering
\begin{tabular}{|ccc|cc|cc|cc|cc|}
\hline
$Z$ & $\tlisa/{\rm~yr}$ & $\nsamp$ & $\nintr$ & $\kintr/\rhocl$ & $\ngttwo$ & $\kgttwo/\rhocl$ & 
$\ngtfive$ & $\kgtfive/\rhocl$ & $\ngtten$ & $\kgtten/\rhocl$ \\
\hline
	0.0001 - 0.02 & 5.0   & 23508 & 1329  &  19.98 & 222  & 3.34 & 72  & 1.08 & 39  & 0.59 \\
	0.005 - 0.02  & 5.0   & 23172 &  917  &  13.99 & 135  & 2.06 & 56  & 0.85 & 29  & 0.44 \\
	0.0001 - 0.02 & 10.0  & 23364 & 1276  &  38.61 & 307  & 9.29 & 104 & 3.15 & 45  & 1.36 \\
	0.005 - 0.02  & 10.0  & 22896 &  924  &  28.53 & 157  & 4.85 & 53  & 1.64 & 30  & 0.93 \\ 
\hline
\end{tabular}
\end{table*}

The peak-power GW frequency in the source frame, $\fgwp$, from a GR in-spiralling binary of
component masses $(\mone,\mtwo)$ and with
instantaneous semi-major-axis $a$ and eccentricity $e$ is given by \citep{Wen_2003}
\beq
\fgwp=\frac{\sqrt{G(\mone+\mtwo)}}{\pi}
	\frac{(1+e)^{1.1954}}{\left[a(1-e^2)\right]^{1.5}}.
\label{eq:gwfreq}
\eeq
The orbital parameters $(a,e)$ decay due to the orbit-averaged leading
gravitational radiation (PN-2.5 term) as (in the source frame) \citep{Peters_1964}
\begin{equation}
\left.
\begin{aligned}
\dot{a} = - \frac{64}{5}\frac{G^3\mone\mtwo(\mone+\mtwo)}{c^5a^3(1-e^2)^{7/2}}
	 \left(1 + \frac{73}{24}e^2 + \frac{37}{96}e^4 \right)\\
\dot{e} = - \frac{304}{15}\frac{G^3\mone\mtwo(\mone+\mtwo)}{c^5a^4(1-e^2)^{5/2}}
	 e\left(1 + \frac{121}{304}e^2\right).
\end{aligned}
\right\rbrace
\label{eq:aedot}
\end{equation}
Note that $\fgwp$ is a certain harmonic, $n_p$, of the Keplerian orbital
frequency, $\fk(=1/2\pi\sqrt{G(\mone+\mtwo)/a^3})$, \ie,
\beq
\fgwp=n_p\fk.
\label{eq:np}
\eeq
Hence, $n_p$ decreases with the binary's orbital evolution (\ie, with decreasing $e$)
such that $n_p\lesssim10$ for $e\lesssim0.7$ and $n_p=2$ for $e=0$.
For low GW frequencies, the frequency time-derivative or `chirp' is given by
\beq
\fdotgwp \approx n_p\dot{f}_{\rm K}
= n_p\frac{48}{5\pi}\frac{(G\mch)^{5/3}}{c^5}(2\pi\fk)^{11/3}F(e)
\label{eq:fgwdot}
\eeq
which can be obtained by utilizing the expression of $\dot{a}$ from Eqn.~\ref{eq:aedot}
(see \eg, \cite{Kremer_2019}). Here
$\mch\equiv(\mone\mtwo)^{3/5}/(\mone+\mtwo)^{1/5}$ is the source-frame chirp mass and
\beq
F(e)\equiv\frac{1}{(1-e^2)^{7/2}}\left(1 + \frac{73}{24}e^2 + \frac{37}{96}e^4 \right)
\eeq
is the `eccentricity correction factor' of Eqn.~\ref{eq:aedot}.

At the peak frequency and from distance $D$,
the characteristic strain (including inclination averaging;
see, \eg, \cite{Peters_1963,Kremer_2019}), $\tilde{h}_c$,
of the GW is given by \citep{Barack_2004,Peters_1963,Kremer_2019}
\beq
\tilde{h}_c^{2} = \frac{2}{3\pi^{4/3}}\frac{G^{5/3}}{c^3}\frac{\mch^{5/3}}{D^2}\frac{1}{\fgwp^{1/3}}
	 \left(\frac{2}{n_p}\right)^{2/3}\frac{g(n_p,e)}{F(e)},
\label{eq:hctid}
\eeq
where $g(n,e)$ is the relative GW power function as given by \citep{Peters_1963}
{\footnotesize
\begin{equation}
\begin{aligned}
 & g(n,e) =  \frac{n^4}{32} \bm\lbrace  \\
 & \left[ J_{n-2}(ne)-2eJ_{n-1}(ne)+\frac{2}{n}J_n(ne)+2eJ_{n+1}(ne)-J_{n+2}(ne) \right]^2 & \\
 & + (1-e^2)\left[ J_{n-2}(ne) - 2J_n(ne) + J_{n+2}(ne) \right]^2 + \frac{4}{3n^2}[J_n(ne)]^2 \bm\rbrace.
\end{aligned}
\label{eq:gne}
\end{equation}
}
$J_n$ is the Bessel function of order $n$ \citep{Press_1992}.

Eqn.~\ref{eq:hctid} can be obtained by combining the general expression for GW
characteristic strain, $\tilde{h}_{c,n}$, for the $n$th harmonic with
source-frame frequency $f^\prime_n=n\fk$
as given by \citep[\eg][]{Barack_2004}
\beq
\tilde{h}_{c,n}^2 = \frac{1}{(\pi D)^2}
\left( \frac{2G}{c^3}\frac{P_n}{\dot{f}^\prime_n} \right)
\label{eq:hcntid}
\eeq
with the average GW power, $P_n$, emitted by the source at the $n$th harmonic
as given by \citep{Peters_1963}
\beq
P_n = \frac{32}{5}\frac{G^4}{c^5}\frac{\mone^2\mtwo^2(\mone+\mtwo)}{a^5}g(n,e).
\label{eq:pn}
\eeq
Using that $\dot{f}^\prime_n\approx n\dot{f}_{\rm K}$ analogously to Eqn.~\ref{eq:fgwdot}
and that $\fk=1/2\pi\sqrt{G(\mone+\mtwo)/a^3}$ (\eg, \cite{Kremer_2019}),
\beq
\tilde{h}_{c,n}^2 = \frac{2}{3\pi^{4/3}}\frac{G^{5/3}}{c^3}\frac{\mch^{5/3}}{D^2}\frac{1}{f^{\prime 1/3}_n}
	 \left(\frac{2}{n}\right)^{2/3}\frac{g(n,e)}{F(e)}
\label{eq:hcntid2}
\eeq
which becomes Eqn.~\ref{eq:hctid} by substituting $f^\prime_n=\fgwp$ and $n=n_p$.

\begin{figure*}
\includegraphics[width=8.5cm,angle=0]{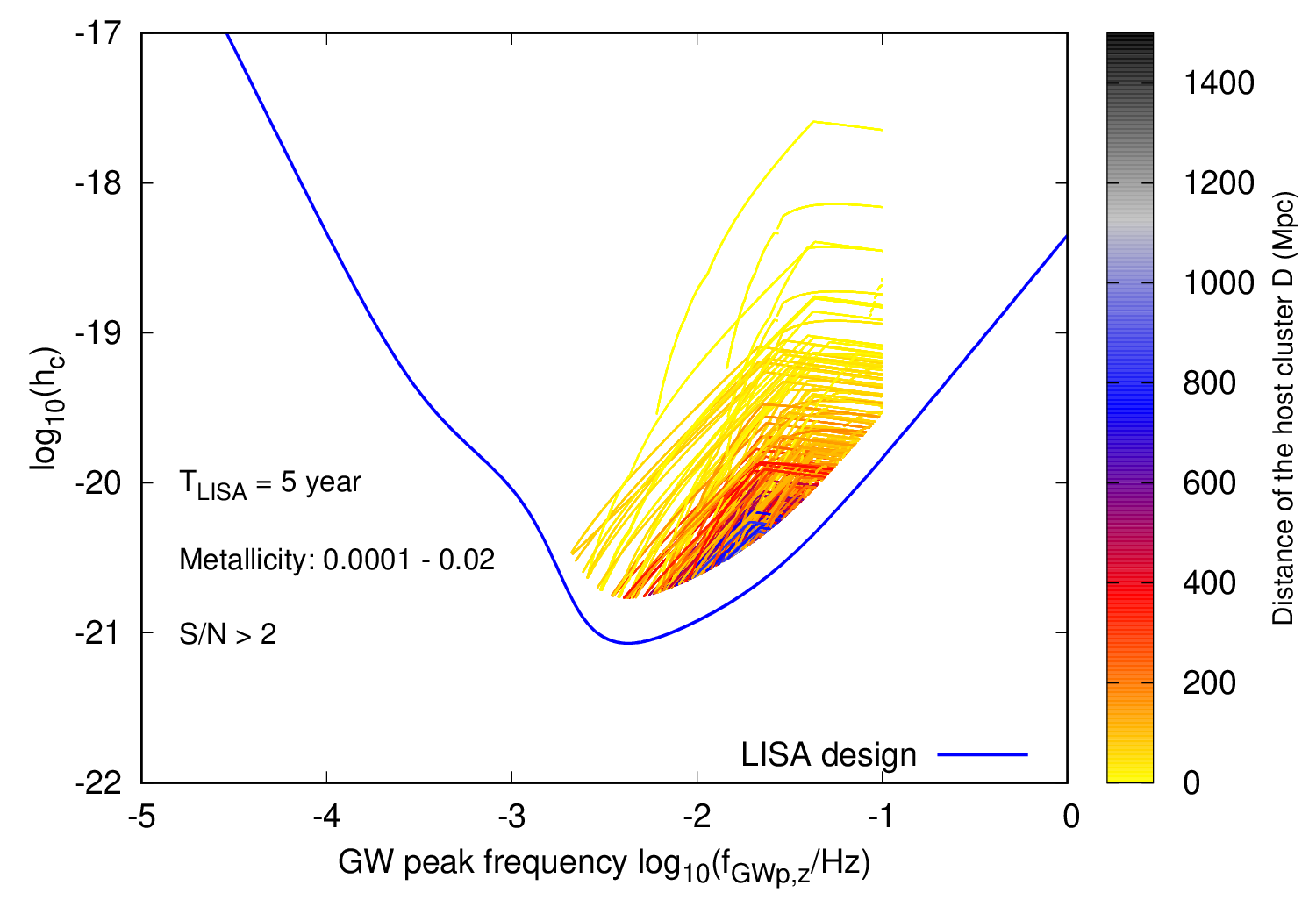}
\includegraphics[width=8.5cm,angle=0]{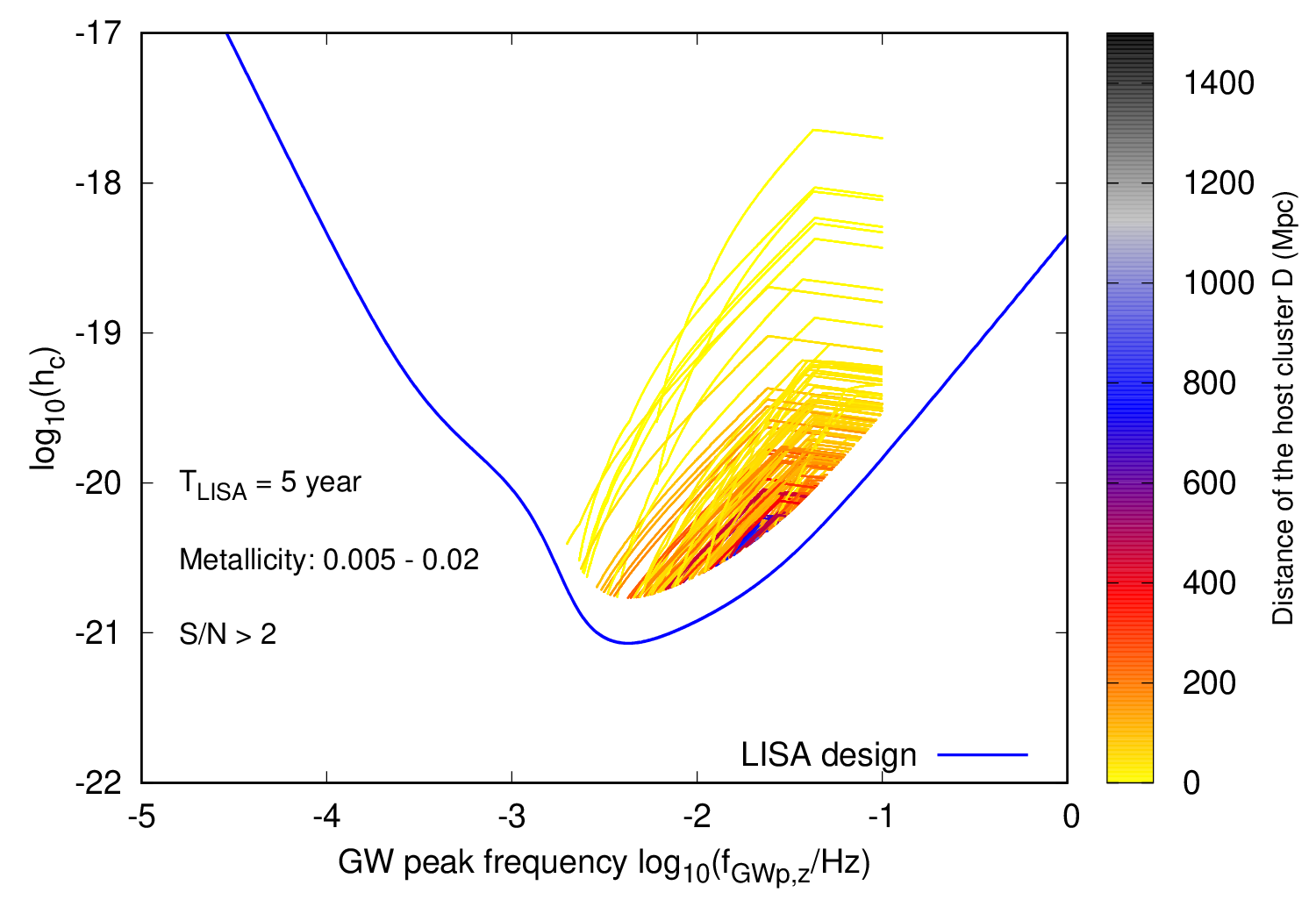}
\caption{BBH inspirals at the present cosmic age (within $\thub\pm0.1$ Gyr),
	over the characteristic LISA frequency range ($10^{-3}{\rm~Hz} - 10^{-1}{\rm~Hz}$),
	from a representative sample of the Local Universe ($D\leq1500$ Mpc)
	as constructed with the computed model clusters (Sec.~\ref{comp}). The inspirals
	are shown (thin lines) in the plane of redshifted GW peak frequency in the detector frame, $\fgwpz$,
	versus GW characteristic strain at this frequency, $\hc$, for those systems
	which have S/N $\geq2$ w.r.t. LISA's design sensitivity curve (thick, blue line).
	The $\hc-\fgwpz$ tracks are colour-coded (colour bar) according to the
	distance of the BBH's host cluster. The left and right panels show
	the outcomes when the metallicity range of the model clusters in the Local-Universe
	sample is taken to be $0.0001-0.02$ and $0.005-0.02$, respectively. A LISA
	mission lifetime of $\tlisa=5{\rm~yr}$ is assumed.
	}
\label{fig:hcplot}
\end{figure*}

\begin{figure*}
\includegraphics[width=8.5cm,angle=0]{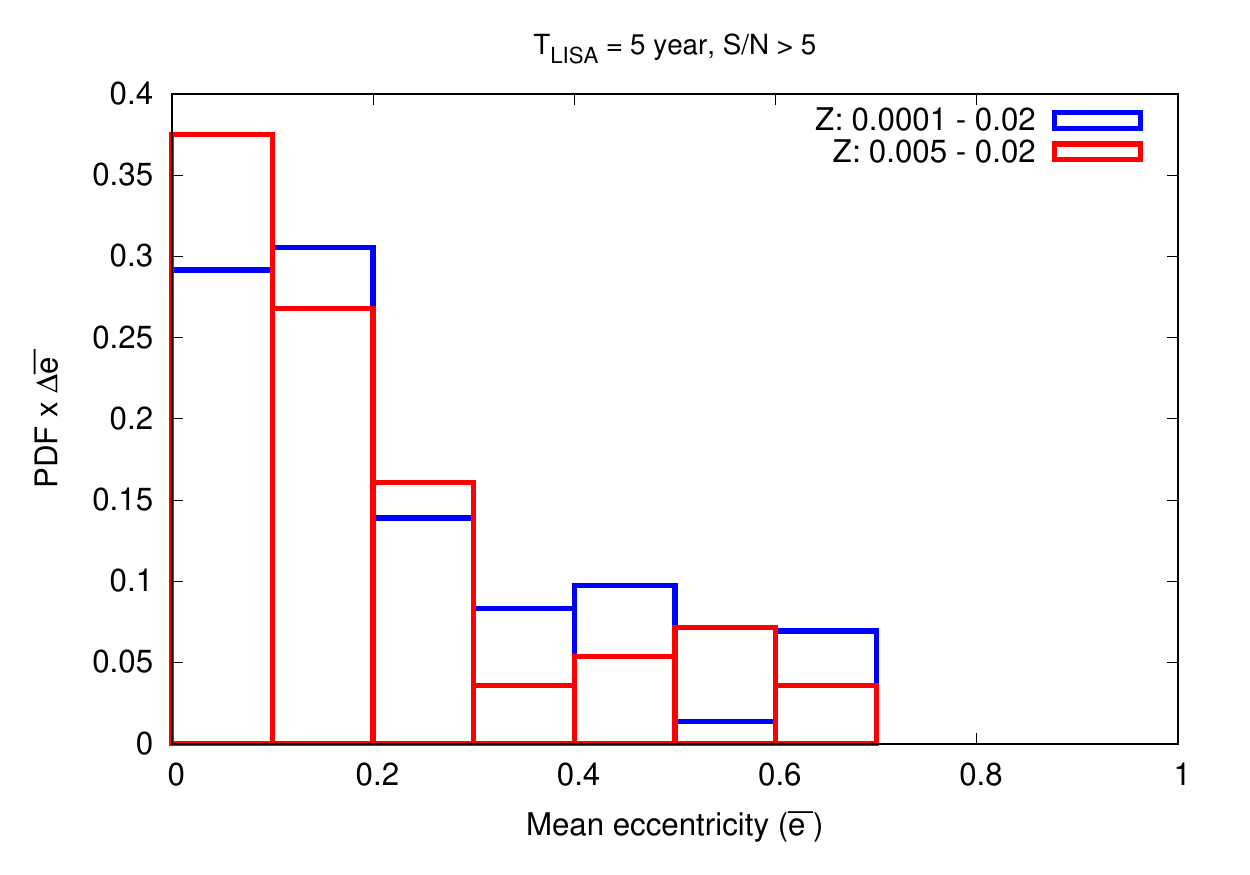}
\includegraphics[width=8.5cm,angle=0]{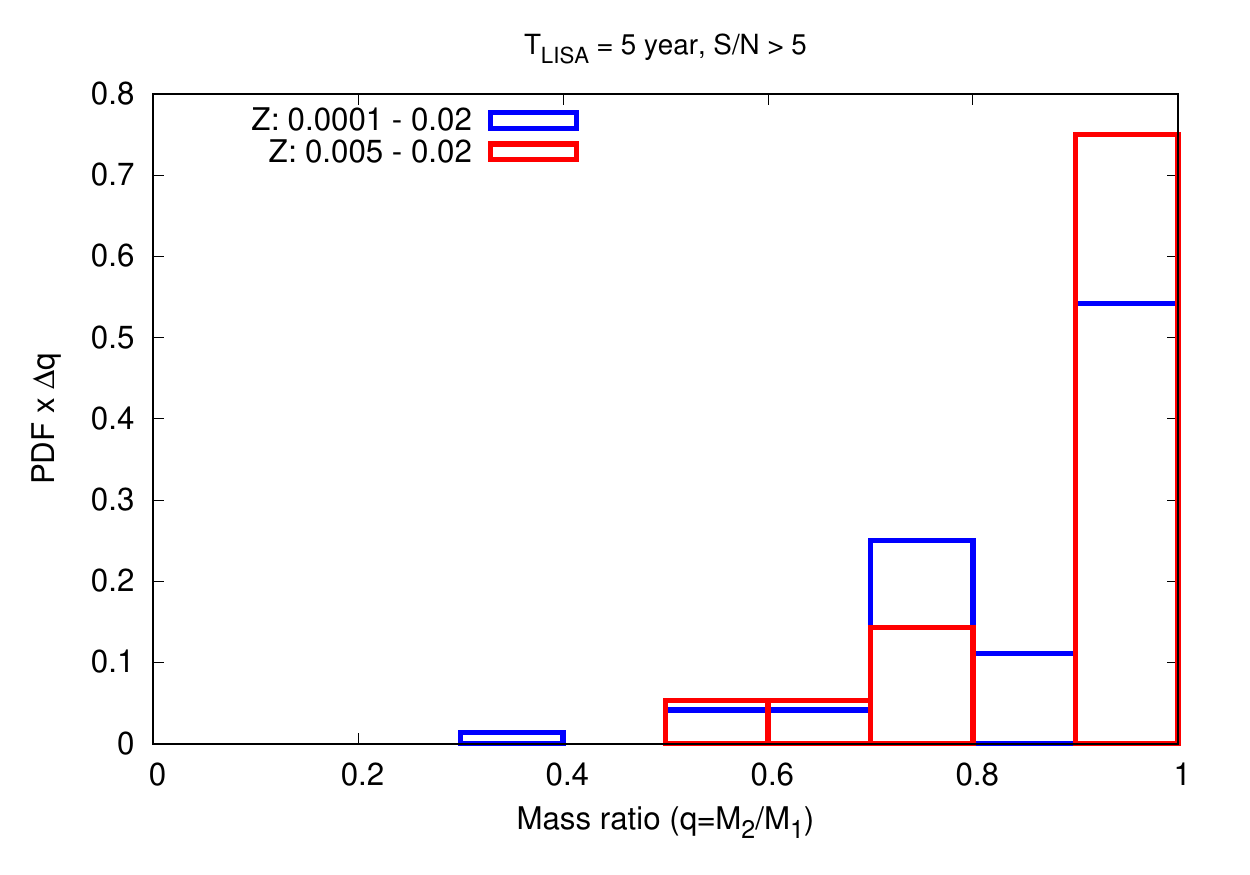}\\
\includegraphics[width=8.5cm,angle=0]{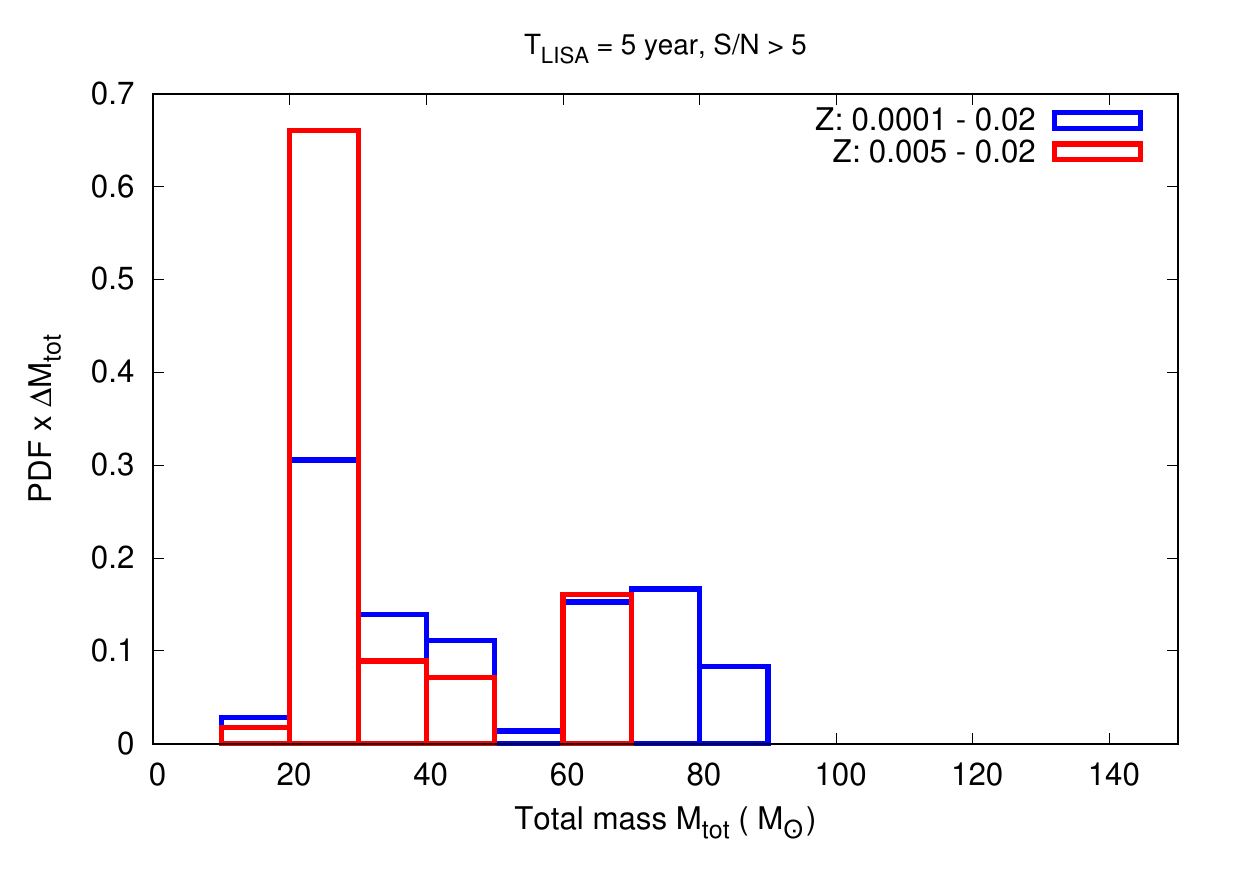}
\includegraphics[width=8.5cm,angle=0]{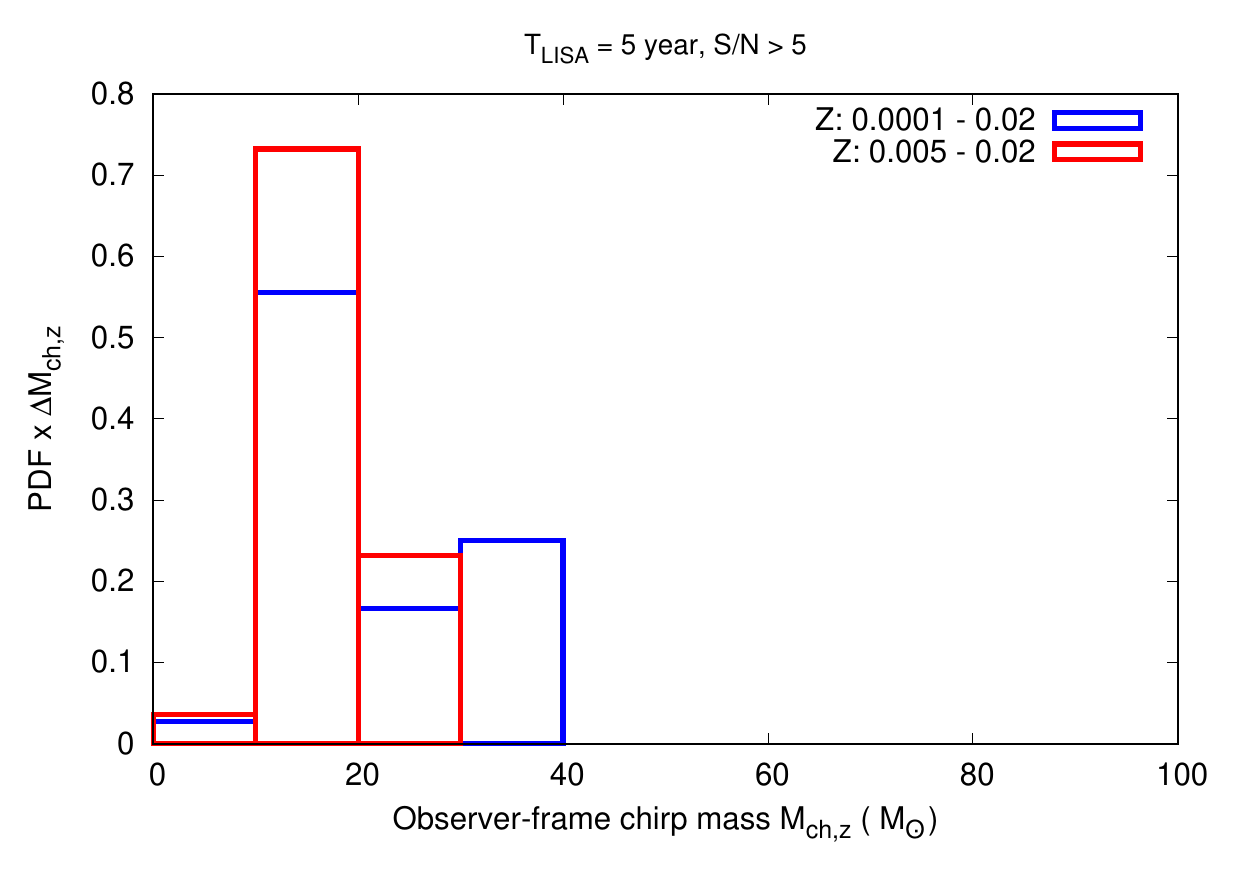}
\caption{Properties of LISA BBH sources, with S/N $>5$, at the present cosmic age from a representative
	Local Universe constructed with the computed model clusters (Sec.~\ref{comp}).
	The top left, top right, bottom left, and bottom right panels respectively show the
	probability distributions of the sources' mean eccentricity (Sec.~\ref{lisasrc}), $\mecc$,
	mass ratio, $q$,
	total mass, $\mtot$, and detector-frame (redshifted) chirp mass, $\mchz$. On each panel,
	the blue- and red-lined histograms correspond to the cluster metallicity ranges
	$0.0001-0.02$ and $0.005-0.02$, respectively. A LISA
	mission lifetime of $\tlisa=5{\rm~yr}$ is assumed.
	}
\label{fig:props}
\end{figure*}

If the redshift at the cluster's distance is $\zd$ then the detector-frame (redshifted) peak GW
frequency, its chirp, and the chirp mass are given by
\begin{equation}
\left.
\begin{aligned}
\fgwpz= & \frac{\fgwp}{(1+\zd)} \\
\fdotgwpz = & \frac{\fdotgwp}{(1+\zd)^2} \\
\mchz= & \mch(1+\zd)
\end{aligned}
\right\rbrace
\label{eq:rshift}
\end{equation}
The detector-frame chirp mass can be obtained by rewriting Eqn.~\ref{eq:fgwdot}
in terms of $\fgwpz$, and $\fdotgwpz$.

In this work, a GW source is considered visible by LISA if, (i), its GW frequency lies
within a relatively narrow range, $10^{-3}{\rm~Hz}\leq\fgwpz\leq 10^{-1}{\rm~Hz}$,
around the instrument's noise floor minimum or `bucket frequency' (at $\sim10^{-2}{\rm~Hz}$;
\cite{eLISA,Robson_2019}). At the same time, (ii), a LISA-visible source should at most be
moderately eccentric, $e\leq0.7$ \citep{Nishizawa_2016,Nishizawa_2017,Chen_2017},
so that it is not `bursty'. Note that a source is effectively lesser detectable
the slower the evolution of $\fgwpz$ ($\fgwp$) is over the LISA mission lifetime, $\tlisa$.
This is taken into account by multiplying a reduction factor to the
GW characteristic strain as given by Eqn.~\ref{eq:hctid}:
\beq
h_c=\kappa\times\tilde{h}_c,
\label{eq:hc}
\eeq
where \citep{Sesana_2005,Willems_2007,Kremer_2019} 
\beq
\kappa = \min\left( \sqrt{\frac{\fdotgwpz}{\fgwpz}\tlisa}, {\rm~~}1 \right).
\label{eq:hcorr}
\eeq
Here, mission lifetimes of $\tlisa=5$ year (planned) and 10 year (optimistic)
are considered.

Finally, (iii), $h_c$ should exceed a signal to noise ratio (hereafter S/N) threshold
for visibility. In this work, LISA sources with S/N $>0$, $\geq2$, $\geq5$, and $\geq10$
are considered; the source count with S/N $>0$ implies the intrinsic count.
The analytical LISA design sensitivity curve (or noise floor) \citep{Robson_2019},
which closely reproduces the instrument's published design sensitivity curve \citep{eLISA} over
the visibility frequency-window considered here, is utilized in determining
S/N for the LISA sources form the sample Local Universe.
In terms of characteristic strain, this sensitivity curve as a
function of (detector-frame) GW frequency, $f$, is given by
\begin{equation}
\left.
\begin{aligned}
\hn(f) = & f^{1/2}\sqrt{\sn(f) + \sgal(f)} \\
\sn(f) = & \frac{\pn(f)}{\respf(f)}.
\end{aligned}
\right\rbrace
\label{eq:hnfull}
\end{equation}
Here, $\pn(f)$ is the power spectral density of the (LISA) instrument noise and
$\respf(f)$ is the sky- and polarization-averaged, two-channel (since
LISA has two independent data channels) signal response function. The added
$\sgal(f)$ is the power spectral density (divided by two due to the two-channel instrument)
of the confusion noise due to unresolved Galactic binaries.

With the analytic fit to the response function, as given by
\beq
\respf(f)=\frac{3}{10}\frac{1}{(1 + 0.6(f/f_\ast)^2)},
\label{eq:resp}
\eeq
\begin{equation}
\begin{aligned}
\sn(f) = & \frac{10}{3L^2}
	\left(P_{\rm OMS}(f) + 2(1+\cos^2(f/f_\ast))\frac{P_{\rm acc}(f)}{(2\pi f)^4}\right)\\
	& \times\left( 1 + \frac{6}{10}\left(\frac{f}{f_\ast}\right)^2 \right).
\end{aligned}
\label{eq:sinst}
\end{equation}
The functions $P_{\rm OMS}(f)$ (single-link optical metrology noise) and $P_{\rm acc}(f)$
(single test mass acceleration noise) are given by (as in ``LISA Strain Curves'' document LISA-LCST-SGS-TN-001)
\begin{equation}
\left.
\begin{aligned}
P_{\rm OMS}(f) = & (1.5\times10^{-11}{\rm~m})^2\left(1+\left(\frac{\rm 2~mHz}{f}\right)^4\right){\rm Hz}^{-1}\\ 
P_{\rm acc}(f) = & (3\times10^{-15}{\rm~m~s^{-2}})^2\left(1+\left(\frac{\rm 0.4~mHz}{f}\right)^2\right)\\
	      &  \times\left(1+\left(\frac{f}{\rm 8~mHz}\right)^4\right){\rm Hz}^{-1}
\end{aligned}
\right\rbrace
\label{eq:lisadoc}
\end{equation}
and the instrument constants are $L=2.5{\rm~Gm}$ (LISA arm length) and $f_\ast=19.09{\rm~mHz}$.
The Galactic confusion noise is given by the fitting function
\beq
\sgal(f)=Af^{-7/3}e^{-f^\alpha+\beta f \sin(Kf)}\left[1+\tanh(\gamma(f_k-f))\right]{\rm Hz}^{-1}
\label{eq:sgal}
\eeq
with $A=9\times10^{-45}$, $\alpha=0.138$, $\beta=-221$, $K=521$, $\gamma=1680$, and
$f_k=0.00113$. See \citep{Robson_2019} and references therein for the derivations of
Eqns.~\ref{eq:resp}, \ref{eq:sinst}, \ref{eq:lisadoc}, and \ref{eq:sgal}.

The resulting LISA design sensitivity curve is shown in the panels of Fig.~\ref{fig:hcplot}
(thick, blue line).
Note that $f_k$ and $\gamma$ in Eqn.~\ref{eq:sgal} vary moderately with observation time
resulting in an increasingly steeper drop-off of $\sgal(f)$. Here, for simplicity, the
4-year values of all the parameters in Eqn.~\ref{eq:sgal}, as stated above \citep{Robson_2019},
are used. Note that over most of the LISA detection frequency range, $\sn$ is the dominant
noise except that $\sgal$ causes the mild `hump' feature in the total sensitivity
curve, as seen in Fig.~\ref{fig:hcplot}.

In this work, for simplicity, S/N is preliminarily taken to be (recalling Eqn.~\ref{eq:hc})
\beq
\sbyn \approx \frac{h_c(\fgwp)}{\hn(\fgwpz)}
      = \frac{\kappa(\fgwpz)\tilde{h}_c(\fgwp)}{\hn(\fgwpz)}
\label{eq:snap}
\eeq
$[\fgwp = (1+\zd)\fgwpz]$ which is evaluated along an in-spiralling orbit using
Eqns.~\ref{eq:gwfreq}, \ref{eq:aedot}, \ref{eq:hctid},
\ref{eq:hcorr}, and \ref{eq:hnfull}-\ref{eq:sgal} (in the practical
computations a lookup table for the LISA design noise strains, generated using Eqns.~\ref{eq:sinst}-\ref{eq:sgal},
is utilized).
A more elaborate expression of S/N is given by summing over all harmonics \citep{OLeary_2009}:
\beq
\sbyn = \sqrt{ \sum_{n=1}^{\infty}\int_{f_{n,0}}^{f_{n,{\tlisa}}}
	\left[\frac{\tilde{h}_{c,n}(f^\prime_n)}{\hn(f_n)}\right]^2d\ln f_n 
	}
\label{eq:snfull}
\eeq
$[f^\prime_n=(1+\zd)f_n]$
\footnote{Alternatively, $\tilde{h}_c$ and $\tilde{h}_{c,n}$ can also
be expressed in terms of quantities in the detector frame using
Eqn.~\ref{eq:rshift} (as, \eg, in \cite{Kremer_2019}). Since, here, the source-frame
quantities are available directly from the computed models, it is natural
to express $\tilde{h}_c$ and $\tilde{h}_{c,n}$ in terms of source-frame GW frequency
and chirp mass and redshift these quantities to the detector frame (Eqn.~\ref{eq:rshift}).}.
Here, $f_n$ is the detector-frame GW frequency of the $n$th harmonic (see above),
and ($f_{n,0}$, $f_{n,{\tlisa}}$) is the GW frequency
of this harmonic at the (start, end) of LISA observation at time ($0$, $\tlisa$).
$\tilde{h}_{c,n}(f^\prime_n)$ is the GW
characteristic strain corresponding to the $n$th harmonic as given by
Eqn.~\ref{eq:hcntid2}.

Since the majority of the LISA-visible binaries have
mild eccentricity (due to condition ii and also GR orbital evolution; see below),
the GW power is sharply peaked at the peak GW frequency $\fgwp$ ($\fgwpz$) \citep{Peters_1963}
which corresponds to the harmonic $n_p\lesssim10$ (see above). In that case, Eqn.~\ref{eq:snfull}
becomes, to the leading order,
\begin{equation}
\begin{aligned}
\sbyn \approx & \sqrt{ \int_0^{\tlisa} \left[\frac{\tilde{h}_c(\fgwp)}{\hn(\fgwpz)}\right]^2
                       \frac{\fdotgwpz}{\fgwpz}dt } \\
      \approx & \sqrt{ \left[\frac{\tilde{h}_c(\fgwp)}{\hn(\fgwpz)}\right]^2
	               \frac{\fdotgwpz}{\fgwpz}\tlisa } \\
	    = & \frac{\kappa(\fgwpz)\tilde{h}_c(\fgwp)}{\hn(\fgwpz)}
\end{aligned}
\label{eq:snap2}
\end{equation}
$[\fgwp = (1+\zd)\fgwpz]$ where the last equality is due to Eqn.~\ref{eq:hcorr}.
In the second approximate relation in
Eqn.~\ref{eq:snap2}, the integrand, to its leading order,
is taken to remain constant at its mean value over $\tlisa$ since its
variation over $\tlisa$ is typically small ($\kappa<1$ for most
sources here; see Fig.~\ref{fig:flim} and the associated
discussions below). Therefore, the approximate
S/N, as given by Eqn.~\ref{eq:snap}, serves as a good approximation for and captures the essential properties
of the full definition, for the LISA sources in this work
\footnote{If, in the presently-considered
LISA frequency range (condition i), $\fgwpz/\fdotgwpz<\tlisa$ (\ie, $\kappa=1$) then
the integral in Eqn.~\ref{eq:snap2} can be subdivided over intervals of $\Delta T<\tlisa$
such that $(\fdotgwpz\Delta T)/\fgwpz\sim1$ and the same approximation can be applied
over each sub-integral, still being consistent with Eqn.~\ref{eq:snap}.
With the approximate S/N evaluated here, marginal sources, whose non-dominant
harmonics would add-up to exceed the S/N threshold, are missed and, consequently,
the present source counts serve as lower limits. However, the underestimation
would be to a small extent since, for the present sources, the GW power spectrum
is sharply peaked at $\fgwp$ due to the sources' mild/small eccentricity (see text).
Similarly, weak but transient (w.r.t $\tlisa$; \ie, $\kappa=1$) sources, whose S/N would
integrate up over $\tlisa$, are also missed. Such underestimation would also
be small since nearly all sources here enter the considered LISA frequency
range with $\kapone<1$ (see Fig.~\ref{fig:flim} and the associated
discussions in the text).}.

The squared value of $\kappa$ (Eqn.~\ref{eq:hcorr}) at the minimum $\fgwpz$ for which
the visibility conditions (i), (ii), and (iii) are simultaneously satisfied
(\ie, at the source's `entry' to the visibility band) is referred to in this work as
the `transience', $\kapone$, of the LISA source. $\kapone\leq1$ is a measure of
how transient the source is over the LISA lifetime: the larger is $\kapone$
the more will the source's $\fgwpz$ and other properties evolve (due to its PN inspiral) over
the mission lifetime. $\kapone=1$ implies that the source evolves in a timescale
$\leq\tlisa$.

In this study, the standard ($\Lambda$CDM) cosmological framework is adopted \citep{Wright_2006}
with the cosmological constants from the latest Planck results
($H_0=67.4\kmps{\rm~Mpc}^{-1}$, $\Omega_{\rm~m}=0.315$, and flat Universe
for which $\thub=13.79{\rm~Gyr}$) \citep{Planck_2018}.

\section{LISA sources from young massive and open stellar clusters}\label{lisasrc}

Fig.~\ref{fig:hcplot} shows examples of BBH inspirals in the LISA band
from a sample Local Universe that are `detected' (\ie, satisfy the visibility
conditions i-iii)
at the present cosmic age with S/N $\geq2$
(\ie, at $\left\lvert \tobs-\thub \right\rvert\leq\delobs\left[=0.1{\rm~Gyr}\right]$),
using the method described in Sec.~\ref{lisacount}. Fig.~\ref{fig:hcplot} shows the detected
inspirals (as dictated by Eqns.~\ref{eq:aedot}, \ref{eq:hctid}, and \ref{eq:hcorr})
in the $h_c-\fgwpz$ plane for the $Z$ ranges 0.0001-0.02 and 0.005-0.02 and
for $\tlisa=5$ year (see Table~\ref{tab_nlisa}).
The design sensitivity curve of LISA \citep{Robson_2019} is shown in the same plane
(the thick, blue line). In the following, unless otherwise stated,
present-day (or present-cosmic-age) LISA sources will imply only those that are detected in the
above sense. Note that depending on the strength of a particular source (given
its distance, mass, and orbital properties), it may be detectable over only a sub-window
within the full detection frequency range $10^{-3}{\rm~Hz}-10^{-1}{\rm~Hz}$ (see Fig.~\ref{fig:hcplot}).

If for the $\tlisa$ mission time the total number of present-day LISA
sources with S/N $\geq s$, from a sample Local Universe comprising
$\nsamp$ clusters, is $\ngts$ then the estimated
number of present-day LISA sources within a $\tlisa$ window, $\kgts$, is
\beq
\kgts = \frac{4}{3}\pi\dmax^3\rhocl\frac{\ngts}{\nsamp}\frac{\tlisa}{2\delobs}, 
\label{eq:nlisa}
\eeq
where $\rhocl$ is the present-day volume density of young clusters in the Local Universe.
Eqn.~\ref{eq:nlisa} can be rewritten as
(for the assumed $\dmax=1500{\rm~Mpc}$; see Sec.~\ref{lisacount}),
\beq
\frac{\kgts}{\rhocl/{\rm~Mpc}^{-3}} = 7.069\frac{\ngts}{\nsamp}
\frac{(\tlisa/{\rm yr})}{(\delobs/{\rm Gyr})}.
\label{eq:nlisasc}
\eeq

Table~\ref{tab_nlisa} shows the $\ngts$ and $\kgts/\rhocl$ values for $s=2$, 5, and 10 for
four Local-Universe samples with $Z$-ranges $0.0001-0.02$ and $0.005-0.02$ and
$\tlisa=5{\rm~yr}$ and 10 yr. Also shown are the intrinsic source counts,
$\nintr$ and $\kintr/\rhocl$, corresponding to S/N $>0$. For each Local Universe,
$\nsamp\approx2.3\times10^4$. By taking half of this $\nsamp$, it is
found that all the source counts also become nearly half, implying that such $\nsamp$
yields statistically convergent counts. 

Fig.~\ref{fig:props} shows the probability distributions (probability density function;
hereafter PDF) of the
properties of LISA BBH sources at the present cosmic age,
that have S/N $>5$, as compiled from Local-Universe samples with $Z$-ranges
$0.0001-0.02$ (blue-lined histogram) and $0.005-0.02$ (red-lined histogram).
All the distributions in Fig.~\ref{fig:props} correspond to $\tlisa=5$ year.
The top-left panel shows the PDF of the `mean eccentricity', $\mecc$,
over the detected GW frequency window. $\mecc$ represents the most likely
eccentricity of the BBH when its GW signal is observed by the detector
and is measured, in this study, by the expression
\beq
\mecc = \frac{(2-\kapone)e_1+\kapone e_2}{2}.
\label{eq:mecc}
\eeq
Here $\kapone$ is the transience of the GW source as
defined in Sec.~\ref{lisacount}. When $\kapone\rightarrow0$ (the source
is nearly invariant over $\tlisa$), $\mecc\rightarrow e_1$, the eccentricity
of the binary at the minimum $\fgwpz$ satisfying the visibility conditions.
When $\kapone=1$ (the source is variable over timescales $\leq\tlisa$),
$\mecc = (e_1+e_2)/2$, midway between the eccentricities,
$e_1$ and $e_2$ respectively ($e_1>e_2$),
at the minimum and maximum $\fgwpz$ satisfying the visibility conditions
(see Fig.~\ref{fig:hcplot}).

The cutoff of the $\mecc$ distribution at $\mecc=0.7$ is simply due to 
the adopted criterion $e\leq0.7$ for visibility by LISA (Sec.~\ref{lisacount}).
Despite the fact that the BBHs' PN inspirals typically start with a high $e$,
the majority of those with $e\leq0.7$ are already well circularized
within the adopted `bucket' frequency range of $10^{-3}{\rm~Hz}-10^{-1}{\rm~Hz}$
(see Fig.~10 of \cite{Banerjee_2020c}; see also \cite{Banerjee_2017b}).
This causes the PDF to increase with decreasing $\mecc$ (top-left
panel of Fig.~\ref{fig:props}). As typical for dynamically-assembled
BBH inspirals \citep{DiCarlo_2019,Banerjee_2020c},
which is the case for the vast majority of inspirals
from the present models, the distribution of the mass-ratio, $q\equiv\mtwo/\mone$ ($\mone\geq\mtwo$),
of the LISA-visible BBHs is strongly biased
towards unity (Fig.~\ref{fig:props}, top-right panel). However, sources as asymmetric as $q<0.4$
is possible for a Local Universe extending to the metal-poorest environments.

The distribution of the total mass, $\mtot$, of the LISA BBH sources
is bimodal (Fig.~\ref{fig:props}, bottom-left panel). The lower mass
peak (spanning over $20\Ms-40\Ms$) is due to the ambience of
lower mass BBH inspirals over the $\lesssim5$ Gyr delay times
considered here (Sec.~\ref{lisacount}); see Fig.~9 of \citep{Banerjee_2020c}.
The higher mass peak, beyond $60\Ms$, appears since despite
the relative rarity of such massive BBH inspirals
they are the brightest GW sources (with highest $\tilde{h}_c$ and $\kappa$).
Note that this bimodal feature appears irrespective of the metallicity range
of the Local Universe. The feature is also mildly present in
the PDF of the detector-frame chirp mass, $\mchz$, of the
LISA BBH sources (Fig.~\ref{fig:props}, bottom-right panel).

\begin{figure*}
\includegraphics[width=12.0cm,angle=0]{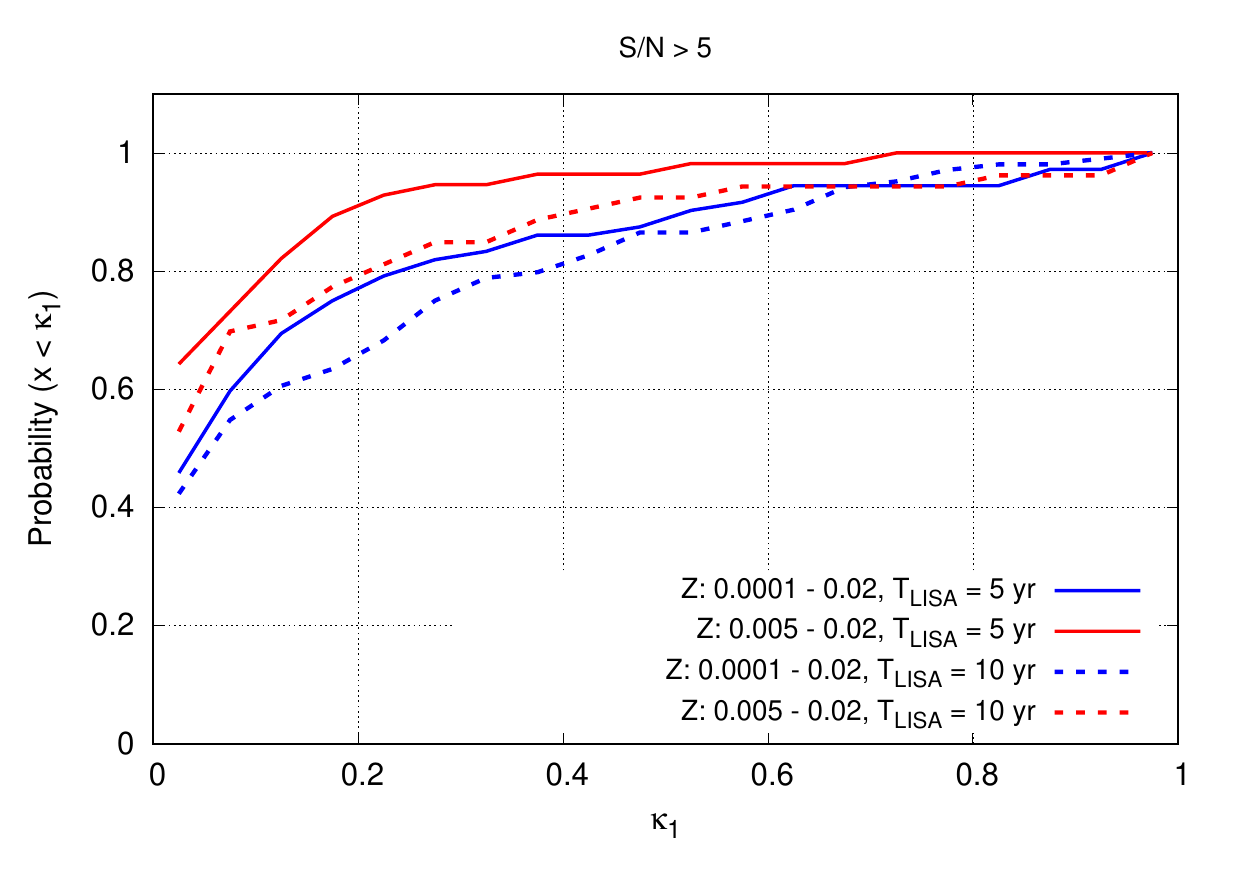}
	\caption{Cumulative probability distribution of the transience, $\kapone$
	(Eqn.~\ref{eq:hcorr}, Sec.~\ref{lisacount}),
	of LISA BBH sources, with S/N $>5$, at the present cosmic age from a representative
	Local Universe constructed with the computed model clusters (Sec.~\ref{comp}).
	The distributions are shown for both metallicity ranges $0.0001-0.02$ and $0.005-0.02$
	and for $\tlisa=5$ and 10 year LISA mission lifetimes, as indicated in the legend.}
\label{fig:flim}
\end{figure*}

Fig.~\ref{fig:flim} shows the cumulative probability distribution of the
transience, $\kapone$ (Sec.~\ref{lisacount}), of the present-day
LISA BBH sources that have S/N $>5$, as compiled from the Local-Universe samples
considered here. Shown are the cumulative PDFs for both metallicity ranges $0.0001-0.02$
(blue lines) and $0.005-0.02$ (red lines)
and for $\tlisa=5$ year (solid lines) and 10 year (dashed lines).
A LISA source would reach merger, \ie, exit the LISA band and become visible
in the LIGO-Virgo band, within twice the LISA mission time if
$\kapone\geq0.5$. According to Fig.~\ref{fig:flim}, with S/N $>5$ and
$\tlisa=5$ year, the fraction of BBHs exhibiting such `LISA-LIGO' visibility
\citep{Sesana_2016} is $\approx15$\% ($\approx5$\%) for the
$0.0001\leq Z \leq0.02$ ($0.005\leq Z \leq0.02$) Local Universe.
With S/N $>5$ and $\tlisa=10$ year, the fraction is $\approx15$\% ($\approx10$\%) for
$0.0001\leq Z \leq0.02$ ($0.005\leq Z \leq0.02$).

With an estimate of the local volume density of YMCs and OCs, $\rhocl$, the values of
$\kgts/\rhocl$ in Table~\ref{tab_nlisa} can be utilized to estimate
the present-day LISA BBH source count. Here a preliminary estimate
of $\rhocl$ is used, which is based on the observed local density of GCs of
$\rho_{\rm GC}\approx2.6{\rm~Mpc}^{-3}$ \citep{vandenBergh_1995,PortegiesZwart_2000}
(taking $h\equiv H_0/[100\kmps]=0.674$). Due to the observed
power-law birth mass function of clusters with index $\approx-2$
\citep{Lada_2003,Gieles_2006a,Gieles_2006b,Larsen_2009} alone, YMCs and OCs
of the mass range considered here ($10^4\Ms-10^5\Ms$) would be
$\approx20$ times more numerous than GCs \citep{Banerjee_2017b},
resulting in $\rhocl\approx52{\rm~Mpc}^{-3}$. The number of LISA
BBH sources, for $\tlisa=5$ year, would accordingly be $\kgttwo\approx 174$,
$\kgtfive\approx 56$, $\kgtten\approx 31$
($\kgttwo\approx 107$, $\kgtfive\approx 44$, $\kgtten\approx 23$)
from the Local Universe with $0.0001\leq Z \leq0.02$ ($0.005\leq Z \leq0.02$).
For $\tlisa=10$ year, $\kgttwo\approx 483$, $\kgtfive\approx 164$, $\kgtten\approx 71$
($\kgttwo\approx 252$, $\kgtfive\approx 85$, $\kgtten\approx 48$)
for $0.0001\leq Z \leq0.02$ ($0.005\leq Z \leq0.02$). For $\tlisa=5$ year,
the intrinsic count for LISA-visible BBHs is $\kintr\approx 1039$ ($\kintr\approx 727$)
from the Local Universe with $0.0001\leq Z \leq0.02$ ($0.005\leq Z \leq0.02$). 
For $\tlisa=10$ year, $\kintr\approx 2008$ ($\kintr\approx 1484$)
for $0.0001\leq Z \leq0.02$ ($0.005\leq Z \leq0.02$).

\section{Summary and discussions}\label{discuss}

This study, for the first time, attempts to assess the potential contribution of YMCs and
OCs, within the Local Universe, in assembling stellar-mass BBHs that are
detectable by LISA as per the instrument's proposed design.
To that end, a suite of state-of-the-art, direct, PN N-body
evolutionary model clusters, incorporating up-to-date 
stellar-evolutionary and remnant-formation models and
observationally-consistent structural properties and stellar
ingredients \citep{Banerjee_2020c}, is utilized (Sec.~\ref{nbcomp}).
The model set allows to explore the cluster mass range of $10^4\Ms-10^5\Ms$
representing the regime where clusters form as YMCs, over the cosmic SFH,
and evolve in long term to become moderately-massive, $\sim{\rm~Gyr}$-old
OCs. In this study, model clusters up to $\approx 5$ Gyr age
(formation redshift $\leq0.5$) are explored,
as typical for intermediate-aged OCs. The BBH inspirals
from them, that would be present at the current cosmic epoch in LISA's most
sensitive GW frequency range of $10^{-3}{\rm~Hz}-10^{-1}{\rm~Hz}$
with eccentricity $<0.7$ and exceeding an S/N threshold (Fig.~\ref{fig:hcplot}),
are tracked (Sec.~\ref{lisacount}). For this purpose, samples
of Local Universe, comprising $\sim 10^4$ clusters (Table~\ref{tab_nlisa})
and having a LISA visibility limit of 1500 Mpc,
are constructed out of the evolutionary cluster models,
following the observed cluster birth mass function and SFH
and adopting the standard cosmological framework (Sec.~\ref{lisacount}).
A sample Local Universe comprises either the full metallicity range
of the cluster models, $0.0001 \leq Z \leq 0.02$ (most of which have $Z\geq0.001$),
implying that the Local Volume well includes LMC-like or sub-LMC metal-poor
environments or only the $0.005 \leq Z \leq 0.02$ models implying
that the Local Volume is made predominantly of metal-rich environments
(Table~\ref{tab_nlisa}).

A drawback of the present approach is that a cluster's
metallicity is completely decoupled from its formation
epoch according to the cosmic SFH. However, this
is not critical since only recent formation redshifts of
$\zf\leq0.5$ are considered. How ambient are metal-poor
environments in the Local Universe is still largely
an open question \citep{Hsyu_2018,Izotov_2018}.
Rather, the two $Z$ ranges considered
here enable exploring the impact of metallicity on LISA source
counts and properties. Indeed, the Local Universe including the metal-poor
clusters typically yields larger, by up to a factor of two,
present-day LISA source counts (Table~\ref{tab_nlisa}). This is due
to the fact that low-$Z$ clusters yield more massive BBH inspirals
(since low-$Z$ stellar progenitors produce more massive BHs
\cite{Belczynski_2010,Banerjee_2020}) so that the present-day LISA BBHs are
biased towards higher masses (Fig.~\ref{fig:props}, bottom panels),
which are also generally brighter GW sources.
In a forthcoming study, metallicity-dependent SFH (\eg, \cite{Madau_2017,Chruslinska_2019})
will be applied in such an exercise.

For both metallicity regimes, the distribution of total mass, $\mtot$, of the present-day
LISA BBHs exhibits a bimodal feature (Fig.~\ref{fig:props}, bottom-left panel;
Sec.~\ref{lisasrc}). For both cases, the present-day LISA BBH sources
are predominantly of similar component masses (mass ratio $q\approx1$) although dissimilar-mass
sources of $q<0.4$ are possible from the metal-poorer Local Universe
(Fig.~\ref{fig:props}, top-right panel). For both type of Local Universe,
the present-day LISA BBH sources are generally eccentric ($e<0.7$),
although they are biased towards being circular (Fig.~\ref{fig:props}, top-left panel;
Sec.~\ref{lisasrc}).

Stellar-mass LISA BBH sources are persistent, with the source properties
varying mildly (as given by their transience, $\kapone$; Sec.~\ref{lisacount})
over the LISA mission lifetime, $\tlisa$, for the
majority of them. However, a small fraction of them would still
exhibit significant evolution as they undergo PN inspiral. 
For the metal-poorer Local Universe, $\approx 15$\% of the present-day LISA BBHs
with S/N $>5$ would show up in the LIGO-Virgo frequency band within
twice the mission lifetime and $<5$\% of the sources would do
so within the mission time, for $\tlisa=5$ year or 10 year
(Fig.~\ref{fig:flim}; Sec.~\ref{lisasrc}). For the metal-richer Local
Universe, the former fraction is $<10$\%. 

Table~\ref{tab_nlisa} shows the estimated number of present-day LISA BBH
sources, $\kgts$, with S/N thresholds $s=2$, 5, and 10, for
both metallicity regimes and for 5 year and 10 year mission
lifetimes. The entries are scaled w.r.t. the present-day volume density, $\rhocl$,
of YMCs and OCs in the Local Universe (see Eqn.~\ref{eq:nlisasc}).
Since such clusters continue to form
and evolve with the cosmic evolution of star formation \citep{Madau_2014},
$\rhocl$ depends on the fraction of stars forming in bound clusters
and the fraction of such clusters surviving the violent birth
environment and conditions \citep{Marks_2012,Marks_2012b,Longmore_2014,Krumholz_2014,Banerjee_2018b,Renaud_2018},
all of which, and hence $\rhocl$, are poorly constrained to date.
By scaling the observed volume density of GCs based on observed cluster
mass function, it can be inferred that YMCs and OCs of $\lesssim 5$
Gyr age in the metal-poorer Local Universe would provide
$\approx 56$ ($\approx 164$) LISA BBH sources
with S/N $>5$, for 5 year (10 year) mission time
(Table~\ref{tab_nlisa}; Sec.~\ref{lisasrc}). For
the metal-richer Local Universe, the corresponding source counts
are $\approx 44$ ($\approx 85$). Therefore, YMCs and OCs
would yield LISA-visible BBHs in about an order of magnitude larger numbers 
than those from GCs \citep{Kremer_2019}. Intrinsically, there would be
$\approx1000$ (700) present-day, LISA-visible BBHs from YMCs and OCs
in the metal-poorer (metal-richer) Local Universe, for $\tlisa=5$ year
(Table~\ref{tab_nlisa}; Sec.~\ref{lisasrc}).
For $\tlisa=10$ year, the intrinsic counts nearly double.

Note that the above estimates of present-day LISA sources still represent
lower limits. The counts can easily be a few factors higher if
the borderline between intermediate-aged OCs and GCs is set at a higher mass (currently,
it is $10^5\Ms$ \cite{Banerjee_2017b}). Also, considering clusters formed
at higher redshifts would add to both the present-day source counts
from a sample Local Universe and the present-day local density of YMCs and
OCs, which would also lead to a few factors boost in the source
counts.

LISA BBH sources in young clusters has been addressed also in other recent
studies \citep{Rastello_2019,DiCarlo_2019}. The eccentricity distribution
of LISA BBH sources, as obtained here,
qualitatively agrees with the trend of the same presented in \citep{DiCarlo_2019}.
But unlike from these authors, the LISA BBH sources here extend to much higher eccentricities,
all the way up to 0.7 (Fig.~\ref{fig:props}, top-left panel).
Note, although, that these authors provide the eccentricity distribution
corresponding to $\fk=10^{-2}$ Hz whereas here the most likely
eccentricity over $10^{-3}{\rm~Hz}\leq\fgwpz\leq 10^{-1}{\rm~Hz}$
(Eqn.~\ref{eq:mecc}) is considered. Furthermore,
in \cite{DiCarlo_2019}, nearly all in-spiralling systems (that merge within a Hubble time) are
dynamically ejected from the clusters whereas here the inspirals take place either inside the clusters
or after getting ejected, the former type being dominant \citep{Banerjee_2017,Anagnostou_2020}.
Finally, the present work considers clusters of much higher mass (by a
few to 100 times) and much longer
evolutionary times than those in \citep{DiCarlo_2019},
yielding BBH inspirals of much broader orbital morphology.
The range and the trend of the eccentricity distribution of LISA BBHs obtained
here are qualitatively similar to those for in-cluster inspirals from computed GC
models \citep{Kremer_2019}, which are a few to 10 times more massive
than the present models but incorporate similar physics ingredients.

In the near future, this line of study will be extended to incorporate
cosmic metallicity evolution and SFH up to high redshifts.
The same methodology can be applied to obtain LIGO-Virgo compact binary merger rates
from YMCs and OCs, which study is underway
(see, \eg, \cite{Santoliquido_2020} for an alternative approach). The present set of
computed model clusters is being extended in mass and density.

\begin{acknowledgments}
SB is thankful to the anonymous referee for constructive comments and useful
suggestions that have helped to improve the manuscript.
SB acknowledges the support from the Deutsche Forschungsgemeinschaft (DFG; German Research Foundation)
through the individual research grant ``The dynamics of stellar-mass black holes in
dense stellar systems and their role in gravitational-wave generation'' (BA 4281/6-1; PI: S. Banerjee).
This work has been benefited by discussions with Sverre Aarseth, Pablo Laguna, Deirdre Shoemaker,
Chris Belczynski, Harald Pfeiffer, Philipp Podsiadlowski, Pau Amaro-Seoane, Xian Chen, Elisa Bortolas,
and Rainer Spurzem. SB acknowledges the generous support and efficient system maintenance of the
computing teams at the AIfA and the HISKP.
\end{acknowledgments}

\bibliography{bibliography/biblio}


\end{document}